\documentclass[aps,prc,superscriptaddress,floatfix,nofootinbib,twocolumn]{revtex4-2}

\usepackage{amsmath,amssymb,amsfonts,physics,graphicx,float,bm}
\newcommand{\Slash}[1]{{\ooalign{\hfil/\hfil\crcr$#1$}}}
\usepackage[setpagesize=false]{hyperref}
\allowdisplaybreaks[4]
\usepackage{subfigure}
\usepackage{mathrsfs}
\usepackage{ulem}
\usepackage{color}
\definecolor{ar}{rgb}{1.0, 0.01, 0.24}
\definecolor{al}{rgb}{0.82, 0.1, 0.26}
\definecolor{ev}{rgb}{0.56, 0.0, 1.0}

\newcommand{\beq}{\begin{eqnarray}}
\newcommand{\eeq}{\end{eqnarray}}

\begin{document}
\title{ 
Analysis of $DD^*$ and $\bar{D}^{(*)}\Xi_{cc}^{(*)}$ molecule by one boson exchange model based on Heavy quark symmetry 
}

\author{Tatsuya Asanuma}
\email{asanuma@hken.phys.nagoya-u.ac.jp}
\affiliation{Department of Physics, Nagoya University, Nagoya 464-8602, Japan}
\author{Yasuhiro Yamaguchi}
\email{yamaguchi@hken.phys.nagoya-u.ac.jp}
\affiliation{Department of Physics, Nagoya University, Nagoya 464-8602, Japan}
\affiliation{Kobayashi-Maskawa Institute for the Origin of Particles and the Universe, Nagoya University, Nagoya, 464-8602, Japan}
\affiliation{Meson Science Laboratory, Cluster for Pioneering Research, RIKEN, Hirosawa, Wako, Saitama 351-0198, Japan}
\author{Masayasu Harada}
\email{harada@hken.phys.nagoya-u.ac.jp}
\affiliation{Department of Physics, Nagoya University, Nagoya 464-8602, Japan}
\affiliation{Kobayashi-Maskawa Institute for the Origin of Particles and the Universe, Nagoya University, Nagoya, 464-8602, Japan}
\affiliation{Advanced Science Research Center, Japan Atomic Energy Agency, Tokai 319-1195, Japan}

\date{\today}

\begin{abstract}
Numerous exotic hadrons with heavy quarks have been reported in the experiments. In such states, symmetries of heavy quarks are considered to play a significant role.
In particular, the superflavor symmetry, or also called the heavy quark anti-diquark symmetry 
is one of the interesting symmetries, which links a quark $Q$ to an anti-diquark $\bar{Q}\bar{Q}$ having the same color representation as $Q$.
In this paper, we study a $\bar{D}\Xi_{cc}$ molecular state as a superflavor partner of the doubly charm tetraquark $T_{cc}$ reported by LHCb recently.
$T_{cc}$ locating slightly below the $DD^*$ threshold is a candidate of the hadronic molecule. 
Thus by replacing the singly charm meson $D^{(*)}$ with the doubly charm baryon $\Xi_{cc}^{(*)}$, superflavor symmetry predicts the existence of the $\bar{D}^{(*)}\Xi_{cc}^{(*)}$ bound state.
We employ the one boson exchange model respecting with the heavy quark spin symmetry where the parameter is obtained to reproduce the $T_{cc}$ binding energy. We apply this model with superflavor symmetry to the $\bar{D}^{(*)}\Xi_{cc}^{(*)}$ molecule and predict a bound state with $I(J^P) = 0(\frac{1}{2}^-)$.
If the pentaquark state corresponding to
$\bar{D}^{(*)} \Xi_{cc}^{(*)}$ molecular state is observed in future experiments as predicted in this work, 
it is more likely that $T_{cc}$ is a $DD^*$ molecular state. 
\end{abstract}

\maketitle

\section{Introduction}
Hadrons are composite particles held together by strong interaction and almost all of them can be classified into baryons ($qqq$) and mesons ($q\bar{q}$). However, in recent years, exotic hadrons that cannot be classified as baryons or mesons have been extensively studied due to many observations of such particles. The existence of exotic hadrons was already 
suggested by Gell-Mann~\cite{gell1964} and Zewig~\cite{Zweig:1964ruk} at the early stage of the quark model. 
In the charm sector, the first candidate of the exotic state, $X(3872)$, was reported in 2003~\cite{choi2003}, triggering further studies of exotic hadrons~\cite{Yamaguchi:2019vea,Brambilla:2019esw,Chen:2022asf}, such as $XYZ$ tetraquarks and $P_c$ pentaquarks. In 2021, LHCb collaboration reported the observation of $T_{cc}^+$ near the $DD^*$ threshold~\cite{lhcb2022}, which is the first doubly charm 
tetraquark ($cc\bar{u}\bar{d}$). The observed state is consistent with an isoscalar state with spin-parity quantum numbers $J^P=1^+$. In the first report~\cite{lhcb2022}, the mass and decay width of the $T_{cc}^+$ were obtained by using the Breit-Wigner parametrisations as $\delta m_{BW}=-273 \pm 61 \pm  5^{+11}_{-14}$ keV and $\Gamma_{BW} = 410 \pm 165 \pm 43^{+18}_{-38}$ keV, respectively, where the mass is measured from the $D^{*+}D^0$ threshold. In Ref.~\cite{LHCb:2021auc}, the model analysis accounting for the relevant thresholds was done, where the pole position was found as $\delta m_{\rm pole} = -360 \pm 40^{+4}_{-0}$ keV, and $\Gamma_{\rm pole} = 48 \pm 2^{+0}_{-14}$ keV.

Doubly heavy tetraquarks $QQ\bar{q}\bar{q}$ including $T_{cc}$ have been studied theoretically in literature~\cite{Chen:2022asf}.
In the 1980s, a genuine exotic doubly heavy tetraquark was discussed in Refs.~\cite{Ader:1981db,Ballot:1983iv,Zouzou:1986qh,Lipkin:1986dw,Heller:1986bt}.
Doubly heavy tetraquarks as compact multiquarks have been studied by various approaches such as the (di)quark model~\cite{Lee:2009rt,Meng:2021yjr,Meng:2020knc}.
The hadronic molecules~\cite{Tornqvist:1993ng} as a two-meson composite state have also been investigated, which have been expected to be realized near thresholds. In fact, the mass of $T_{cc}$ reported by LHCb 
is just below the $DD^*$ threshold. There have been many studies of doubly heavy tetraquarks as hadron molecule by using 
the meson exchange model ~\cite{Manohar:1992nd,Ohkoda:2012hv,li2013,Cheng:2022qcm}, the effective field theory~\cite{Pan:2022whr}, and so on.
Lattice QCD simulation has also been a useful approach to investigate the exotic hadrons from QCD, where $T_{cc}$ tetraquarks have been studied in Refs.~\cite{Ikeda:2013vwa,Bicudo:2015vta,Junnarkar:2018twb,Padmanath:2022cvl,Bicudo:2022cqi,Lyu:2023xro}.

Observations of exotic hadrons containing heavy quarks indicate that the understanding of these states is greatly aided by the concept of heavy quark symmetry. Specifically, the heavy quark spin interaction is suppressed by a factor of $\mathcal{O}(1/{m_Q})$, which 
induces Heavy Quark Spin Symmetry (HQSS) (see, for a review Ref.~\cite{manohar2000}). 
This symmetry leads to 
the heavy quark spin multiplet 
structure. In a case of heavy-light mesons, mass differences of a pseudoscalar meson with spin $0$ and a vector meson with spin $1$ are (approximately) suppressed. 
Actually in the charm quark sector, the mass difference between $D^*(2008)$ and $D(1867)$ is about $140$ MeV, which is as small as $m_{\pi}$. 

Additionally, there exists an interesting symmetry which is the superflavor symmetry, or also called the heavy quark anti-diquark symmetry. 
This symmetry arises from the replacement of a heavy quark $Q$ with a heavy anti-diquark $\bar{Q} \bar{Q}$ having the same color representation $3_c$ as $Q$~\cite{georgi1990,savage1990,Ma:2017nik}, 
indicating that 
$\bar{D} (\bar{Q}q)$ and $\Xi_{cc} (QQq)$ is related. 
Thus, the superflavor symmetry predicts an existence of multi-heavy hadronic states corresponding to the $\bar{D}^{(*)}$ meson molecules by replacing $\bar{D}^{(*)}$ mesons with $\Xi_{cc}^{(*)}$ baryons.

In this paper, we study a $\bar{D}^{(*)}\Xi_{cc}^{(*)}$ molecular state as a superflavor partner of the 
$T_{cc}$ tetraquark assumed to be a $D^{(*)}D^{(*)}$ molecule.
We firstly construct a model to describe $T_{cc}$ reported by LHCb as the
${D}^{(*)} {D}^{(*)}$ 
molecule with $I(J^P)=0(1^+)$. The one boson exchange model (OBEM) is employed as for the interactions between $D^{(*)}$ 
mesons, where a cutoff parameter $\Lambda$ in OBEM 
is determined to reproduce the experimental data of $T_{cc}$. As the binding energy of $T_{cc}$ referred in this study, we employ the Breit-Wigner mass $\delta m_{\rm BW}$ given in the first report of LHCb. However In this study, we employ the Breit-Wigner mass $\delta m_{\rm BW}$ as the binding energy of $T_{cc}$. As far as the bound-state problem is considered, the qualitative results do not change if the pole mass $\delta m_{\rm pole}$ is employed.
there are no qualitative difference between cases employing $\delta m_{\rm BW}$ and $\delta m_{\rm pole}$. 
We conduct a coupled-channel analysis of $DD^*$ and $D^*D^*$ respecting HQSS due to the small mass difference of $m_{D^*} - m_D \simeq 140$ MeV. 
We expand this model into a $\bar{D}^{(*)} \Xi_{cc}^{(*)}$ molecular state by using superflavor symmetry. 
We obtain effective Lagrangians for $\Xi_{cc}^{(*)}$, where the coupling constants are estimated from those of the $D^{(*)}$ mesons based on superflavor symmetry.
The coupled-channel analysis including the possible $\bar{D}^{(*)}\Xi_{cc}^{(*)}$ channels are performed, respecting HQSS due to the small mass difference of $m_{\Xi^*_{cc}}$ and $m_{\Xi_{cc}}$, about $105$ MeV. 
We study a $\bar{D}^{(*)} \Xi_{cc}^{(*)}$ bound state as a superflavor partner of the $T_{cc}$ state being the $D^{(*)}D^{(*)}$ molecule. 
We also discuss the breaking effect of superflavor symmetry by changing the value of the coupling constants.

The paper is organized as follows. In Sec.~\ref{sec:DD*molecule}, we study the $D^{(*)}D^{(*)}$ molecule. We construct the OBEM to describe the reported $T_{cc}$. In Sec.~\ref{sec:DXiccmolecule}, the OBEM is applied to the $\bar{D}^{(*)}\Xi_{cc}^{(*)}$ molecule. We study a possible existence of the $\bar{D}^{(*)}\Xi_{cc}^{(*)}$ bound states as a  superflavor partner of the $D^{(*)}D^{(*)}$ molecule. Finally Sec.~\ref{sec:summary} is devoted to the summary.

\section{$DD^*$ molecule}
\label{sec:DD*molecule}

We perform a coupled-channel analysis of $D^{(*)}D^{(*)}$ molecular state with $I(J^P) = 0(1^+)$, 
associated with $T_{cc}$ reported by LHCb.
Respecting HQSS due to the small mass difference $m_{D^*} - m_D \simeq 140 $ MeV, 
we consider that $DD^*-D^*D^*$ channel coupling is important.
We also include $S$-wave and $D$-wave channels mixed by the tensor force which is important for deutron to make the bound state~\cite{Ikeda:2010aq,yamaguchi2011}.
Thus, we consider the following coupled-channel
\begin{align}
\begin{pmatrix}
[DD^*]_-(^3S_1)\\
[DD^*]_-(^3D_1)\\
D^*D^*(^3S_1)\\
D^*D^*(^3D_1)
\end{pmatrix}   
,
\label{eq:DD*channels}
\end{align}
for $I(J^P)=0(1^+)$~\cite{Ohkoda:2012hv}
where
$[DD^*]_- = \frac{1}{\sqrt{2}}(DD^* - D^*D)$.

In order to derive the one boson exchange potential (OBEP) 
for the $D^{(*)}D^{(*)}$ state, 
we employ 
the effective Lagrangians for $D^{(*)}$ mesons and light mesons based on the HQSS. 
The heavy meson doublet field $H$ is introduced as 
a linear combination of pseudo-scalar meson $P$ and vector meson $P^{*\mu}$ degenerating in the heavy quark limit (HQ limit) as follows~\cite{choi2003},

\begin{align}
    H &= \frac{1 + \Slash{v}}{2} [\gamma_\mu P^{*\mu} - \gamma^5 P] 
    \, , \\
    \bar{H} &= \gamma^0 H^{\dagger} \gamma^0 \notag \\
            &= [\gamma_\mu P^{*\mu \dagger} + \gamma^5 P^\dagger] \frac{1 + \Slash{v}}{2} 
           \, ,
\end{align}
where the isospin index is omitted.
$v^\mu$ is the four velocity of the heavy meson.
Then, we obtain 
the effective Lagrangians of the $H$ fields with the Nambu-Goldstone bosons
in the leading order of the $1/m_Q$ expansion, 
written by
\begin{equation}
  \begin{aligned}
 \mathcal{L}_{H H \mathcal{M}} =& g \operatorname{Tr}\left[H_{b}\gamma_{\mu}\gamma_{5}A_{b a}^{\mu} \bar{H}_{a}\right]\\
 =& 2g P_{a}^{* \mu \dagger} P_b A_{\mu ba} + 2g P_{a}^{\dagger} P^{*\mu}_b A_{\mu ba} \\
 &- 2ig \varepsilon^{\mu \nu \rho \sigma} v_{\mu} P_{\nu a}^{*\dagger} P^{*}_{\rho b} A_{\sigma ba} ,
\end{aligned}
\end{equation}
which is invariant under the HQSS and $SU(3)$ flavor symmetry.
The axial current $A^\mu$ and the meson field are given by
\begin{align}    
A^{\mu} &= \frac{i}{2}\left(\xi^{\dagger} \partial^{\mu} \xi-\xi \partial^{\mu} \xi^{\dagger}\right) \, , \\
\xi &= e^{\frac{i 
    \hat{\pi} 
        }{\sqrt{2}f_{\pi}}} = e^{i \frac{\pi^a \tau^a}{2 f_{\pi}}} \, ,
\end{align}
where
\begin{align}
    \hat{\pi}
    &= \left[\begin{array}{ccc}
\frac{1}{\sqrt{2}} \pi^{0}+\frac{\eta}{\sqrt{6}} & \pi^{+} & K^{+} \\
\pi^{-} & -\frac{1}{\sqrt{2}} \pi^{0}+\frac{\eta}{\sqrt{6}} & K^{0} \\
K^{-} & \bar{K}^{0} & -\frac{2}{\sqrt{6}} \eta
\end{array}\right] \, ,
\end{align}
and $f_\pi = 92.4$\,MeV is the pion decay constant. 
We obtain the odd number of $\pi$ from the expansion of $A^\mu$ and the even number of $\pi$ from the expansion of $V^\mu$. 

The interaction Lagrangians for the scalar meson $(\sigma)$ 
and the vector mesons 
$\rho$ and $\omega$ 
are given by~\cite{casalbuoni1993}
\begin{equation}
    \begin{aligned}
\mathcal{L}_{H H \sigma} &= g_{\sigma} \operatorname{Tr}\left[H_{a}\sigma\bar{H}_{a}\right]\\
&= -2g_{\sigma} P_{b} P_{b}^{\dagger} \sigma + 2 g_{\sigma} P_{b}^{*}P_{b}^{* \dagger} \sigma \, .
\end{aligned}
\end{equation}
and 
\begin{equation}
    \begin{aligned}
\mathcal{L}_{H H V}
= &\,  
     \beta \operatorname{Tr}\left[H_{b} v_{\mu}\left(V_{b a}^{\mu}-\rho_{b a}^{\mu}\right) \bar{H}_{a} \right]\\ &+ \lambda \operatorname{Tr}\left[H_{b} \sigma_{\mu \nu} F^{\mu \nu}(\rho)_{b a} \bar{H}_{a}\right] \notag \\
\simeq & \,  -\sqrt{2}\beta g_V \left( P^{b\dagger}P_a - P^{*b\dagger}P^{*}_a \right)v_{\mu} \hat{\rho}^{\mu}_{ba} \\
&+ i2\sqrt{2}\lambda g_V P^{*b\dagger}_\mu P^{a*}_\nu (\partial^\mu \hat{\rho}^\nu_{ba} - \partial^\nu \hat{\rho}^\mu_{ba}) \notag \\
& - 2\sqrt{2}\lambda g_V \varepsilon_{\mu \nu \rho \sigma} v^{\mu} (P^{b*\nu \dagger}P^a
+P^{b\dagger}P^{a*\nu})\partial^\rho \hat{\rho}^{\sigma}_{ba}
\end{aligned}
\end{equation}
where 
\begin{align}
V^{\mu} &= \frac{i}{2}\left(\xi^{\dagger} \partial^{\mu} \xi+\xi \partial^{\mu} \xi^{\dagger}\right) \\
F_{\mu \nu}(\rho) &= \partial_{\mu} \rho_{\nu}-\partial_{\nu} \rho_{\mu}-\left[\rho_{\mu}, \rho_{\nu}\right] \\
\rho_{\mu} &= -\frac{g_v}{\sqrt{2}} \hat{\rho}_{\mu} = -\frac{g_v}{2} \rho^a_\mu \tau^a \\
\hat{\rho}^{\mu} &= \left(\begin{array}{ccc}
\frac{\rho^{0}}{\sqrt{2}}+\frac{\omega}{\sqrt{2}} & \rho^{+} & K^{*+} \\
\rho^{-} & -\frac{\rho^{0}}{\sqrt{2}}+\frac{\omega}{\sqrt{2}} & K^{* 0} \\
K^{*-} & \bar{K}^{* 0} & \phi 
\end{array}\right)^{\mu} \, .
\end{align}

We list the coupling constants in Table~\ref{table:coupling_constant}. $g = 0.59$ for the $\pi$ interaction  
is determined from the decay width of $D^{*-} \rightarrow D^- \pi^0$~\cite{ahmed2001}. 
For the vector meson interactions, 
we use $\beta \simeq 0.9 $ and $\lambda=0.56$ GeV$^{-1}$ from the form factor approach in $B$ meson decay~\cite{liu2019}, which is supported in the Lattice QCD calculation~\cite{isola2003}.
For the coupling constant of the $\sigma$ meson, $g_\sigma$, different values have been used in the literature.
In Refs.~\cite{Riska:1999fn,chen2017},
$g_\sigma$ is estimated as 
$g_\sigma=\frac{g_{\sigma NN}}{3}=3.4$ 
with the sigma-nucleon coupling constant $g_{\sigma NN}$. The factor of $1/3$ is chosen by assuming that the $\sigma$ meson couples to light quarks inside hadrons.
The nucleon contains three light quarks, while the $D^{(\ast)}$ meson does one light quark and one anti heavy quark. 
Alternatively, in Refs.\cite{Liu:2008xz,Li:2012ss}, the small value of the sigma coupling is obtained as $g_\sigma=g_\pi/2\sqrt{6}=0.76$, where $g_\pi\sim 3.73$ is given by the Goldberger-Treiman relation and $m_{D_s^\ast}-m_{D_s}\sim 349$ MeV~\cite{Bardeen:2003kt}. In this study, we compare the results with large and small values of $g_\sigma$, denoted by 
$g_\sigma^{(1)}=3.4$ and $g_\sigma^{(2)}=0.76$.

For 
the hadrons masses, 
we employ the isospin average of the masses listed in the PDG~\cite{ParticleDataGroup:2022pth}. 

\begin{table}[htbp]
    \centering
    \caption{Coupling constants of the effective Lagrangians.}   
    \label{table:coupling_constant}
    \begin{tabular}{lcc} 
        \hline\hline
          Coupling const.  & Values & Refs \\ \hline
        $g$ & 0.59  &     \cite{ahmed2001}\\
        $g_\sigma^{(1)}$  & 3.4  & \cite{chen2017}\\
        $g_\sigma^{(2)}$  & 0.76  & \cite{chen2017}\\    
        $g_v$ & 5.9  & \cite{casalbuoni1993}\\
        $\beta$ & 0.9  & \cite{liu2019}\\
        $\lambda$ & 0.56 \, $\mathrm{GeV^{-1}}$ & \cite{isola2003}\\
         \hline\hline         
    \end{tabular}
    \label{tab:my_label}
\end{table}

From those Lagrangians, we derive the OBEP in momentum space using the formula 
\begin{align}
    V(\bm{q}) = \frac{i\mathcal{M}}{\sqrt{\Pi_i 2m_i \Pi_f 2m_f}}    \, ,
\end{align}
where $\mathcal{M}$ 
is the $t$-channel scattering amplitude of the meson exchange process,
and $m_i$ and $m_f$ are masses of initial and final hadrons, respectively. 
To account for the finite size of hadrons, we introduce a monopole-type form factor 
\begin{align}
 F(\bm{q}) = \frac{\Lambda^2 - \mu_{eff}^2}{\Lambda^2 + \bm{q}^2} \, ,   
\end{align}
 where the cutoff parameter $\Lambda$ 
is a free parameter in our model. In this work, we determine the value to reproduce the $T_{cc}$ experimental data 
($\delta_{m_{BW}} \simeq 0.273$ MeV).
 $\mu_{eff}$ is the effective mass defined as $ \mu_{eff}^2 = m_{ex}^2 -(q^0)^2$, where $m_{ex}$ is the mass of the exchanged meson and $q^0$ 
is the zeroth component of the momentum given by $q^0 \simeq \frac{m_2^2-m_1^2+m_3^2-m_4^2}{2\left(m_3+m_4\right)}$ ~\cite{Li:2012ss}. 
 In the case for
 $\mu_{eff}^2 < 0$, the potential has an imaginary part implying a decay into the $DD\pi$ channel~\cite{Suzuki:2005ha,Yamaguchi:2019vea,Cheng:2022qcm}. Since we focus on the bound state problem in this study, 
 we take only the real part of the potential.
The OBEP in coordinate space 
is obtained by the Fourier transformation
\begin{align}
 V(r) = \frac{1}{(2\pi)^3}\int d^3 q V(\bm{q}) [F(\bm{q})]^2   \, .
\end{align}
We note that the contact term given by the delta function in the central part of the potential is subtracted in this paper.

\begin{table}[htbp]
    \centering
 \caption{
 Bound state properties of the $D^{(\ast)}D^{(\ast)}$ molecule with various cutoff $\Lambda$ and fixed $g_\sigma=g_\sigma^{(1)}=3.4$.
            $\mathrm{B_{in}}$ is a binding energy measured from the $DD^\ast$ threshold. $\Lambda$ and $\mathrm{B_{in}}$ are given in units of MeV. $P$ is probability of each channels. 
            $\sqrt{\langle r^2 \rangle}$ is a root mean square radius in unit of fm.
            The results with $\Lambda=1182$ MeV reproduce the binding energy of $T_{cc}$ in LHCb~\cite{lhcb2022}.
            }   
        \label{tab:results_DDast}    
    \begin{ruledtabular}
    \begin{tabular}{lccc}    
    $\Lambda$ [MeV] & 1160 & 1182 & 1200 \\
    \hline 
    $\mathrm{B_{in}}$ [MeV] & 0.074 & 0.273
        & 0.549 \\
    $P_{[DD^*]_{-}}(^3S_1)$ & 0.992 & 0.987 & 0.983 \\
    $P_{[DD^*]_{-}}(^3D_1)$ & 0.00545 & 0.00840 & 0.0103\\
    $P_{D^*D^*}(^3S_1)$ & 0.00159 & 0.00348 & 0.00541\\
    $P_{D^*D^*}(^3D_1)$ & 0.000542 & 0.00106 & 0.00151 \\
    $\sqrt{\langle r^2 \rangle}$ [fm]& 11.33 & 6.42 & 4.70\\ 
    \end{tabular}
    \end{ruledtabular}
\end{table}

\begin{figure}[htbp] 
\centering
\includegraphics[width=0.9\linewidth,clip]{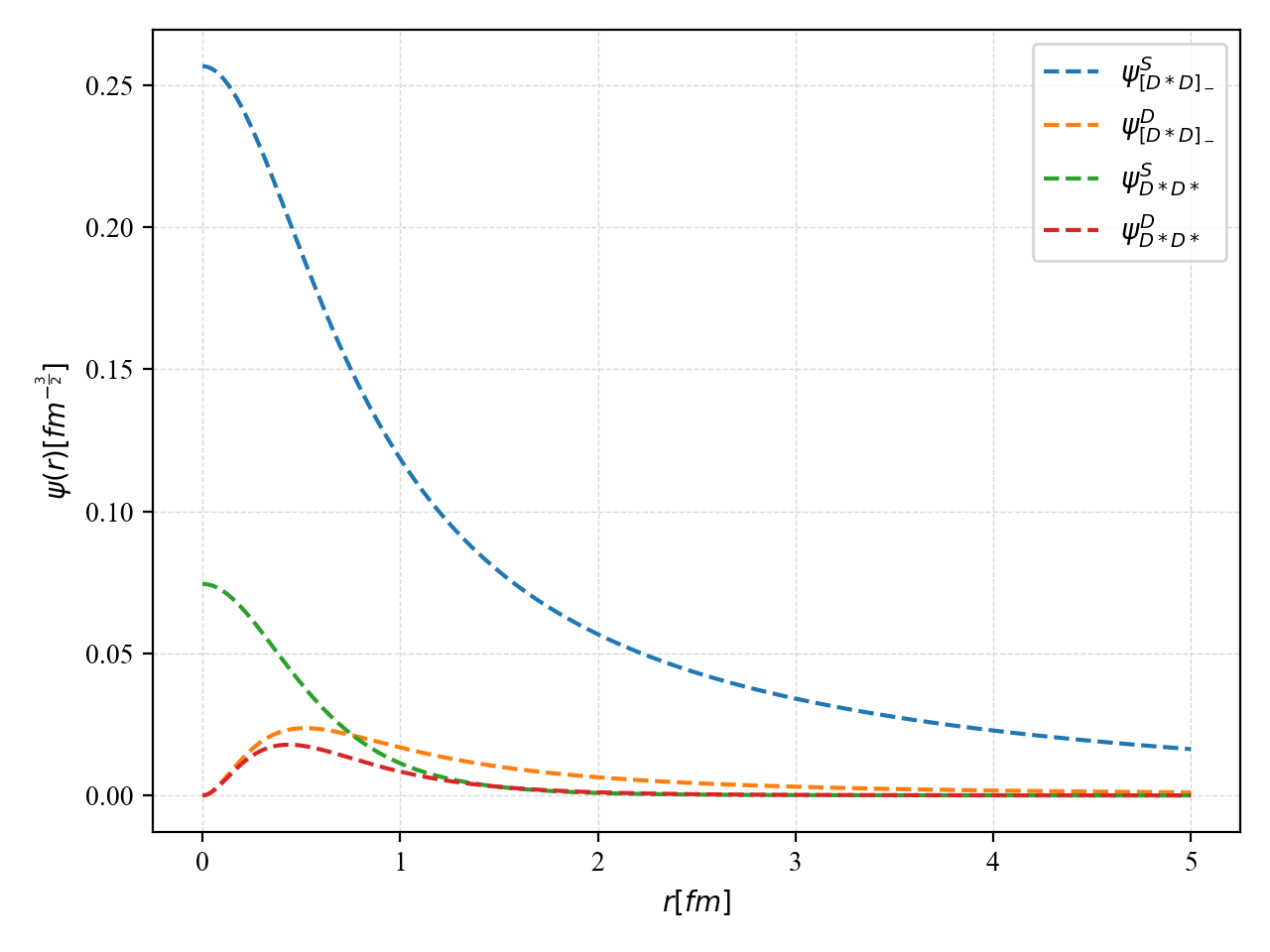}
\caption{Wave functions of the $D^{(*)}D^{(*)}$ molecule with $g_\sigma = g_\sigma^{(1)}= 3.4$. The value $\psi(r)$ is given in unit of $[\rm{fm}^{-\frac{3}{2}}]$.}
\label{fig:wave_func_DDast_gs3.4}
\end{figure}

\begin{figure}[htbp]
\centering
\includegraphics[width=\linewidth,clip]{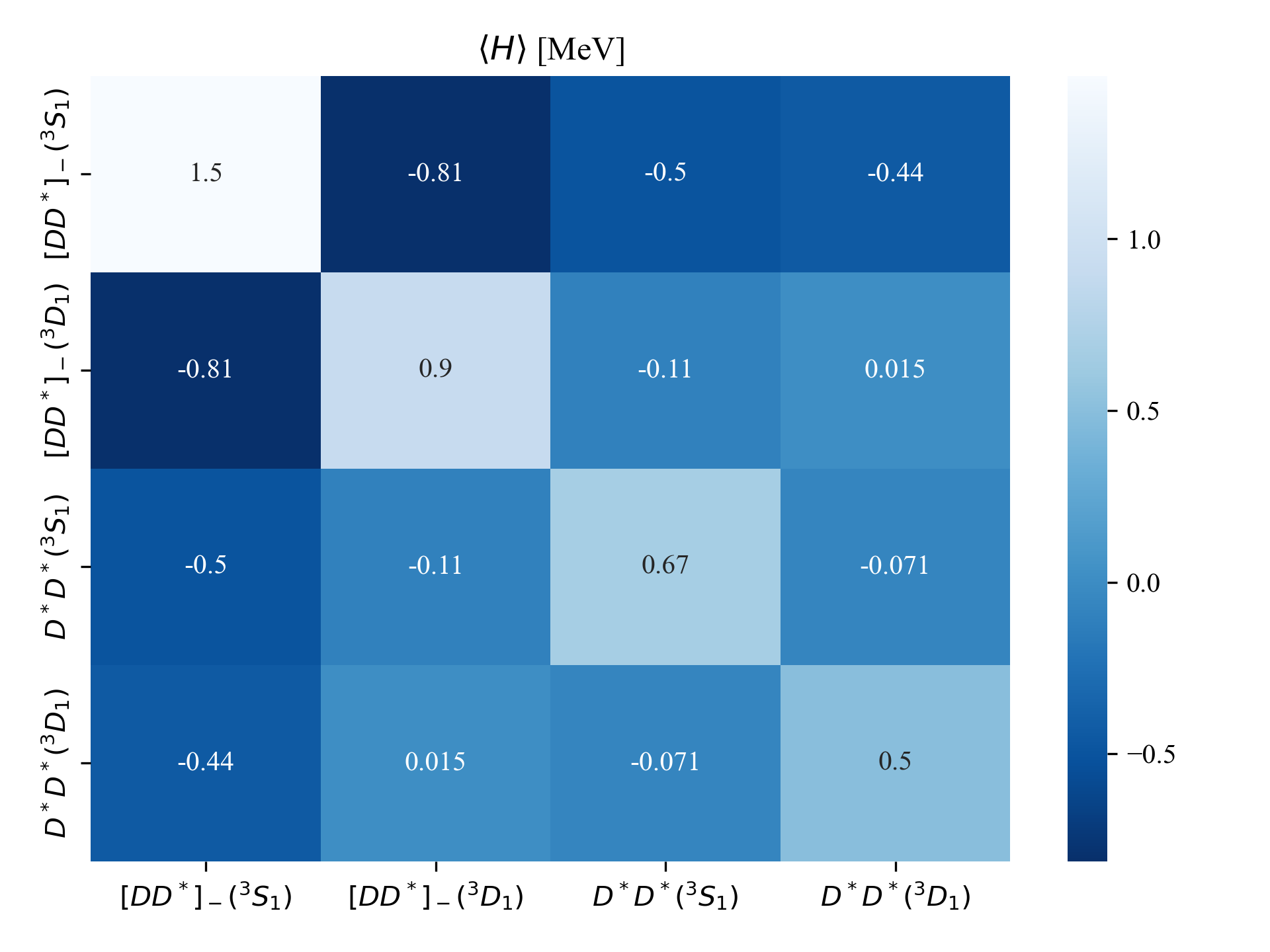}
\caption{
Energy expectation values $\langle H \rangle$ [MeV] of the $D^{(*)}D^{(*)}$ molecule with $g_\sigma = g_\sigma^{(1)}= 3.4$. Positive (negative) values indicate their repulsive (attractive) contribution. The values are given in units of MeV.
}
    \label{fig:expvalue_DDast_gs3.4}
\end{figure}

We solve the coupled-channel Schr\"odinger equations with the obtained potentials using the Gaussian expansion method~\cite{hiyama2003}, and study a bound state of the $DD^\ast$ molecule.
By using appropriate value of the cutoff $\Lambda$, 
we obtain a bound state. 

First we employ the large value of the sigma coupling $g_\sigma=g_\sigma^{(1)}=3.4$.
Numerical results of bound states with various values of $\Lambda$ are summarized in Table~\ref{tab:results_DDast}.
The table presents the values of $\Lambda$, the binding energy, the probability of each channel, and the root mean square (RMS) radius. 
By setting $\Lambda \simeq 1182$ MeV, we reproduce the $T_{cc}$ binding energy in the experimental data.
The radial wave function and the matrix of the Hamiltonian expectation value 
with this cutoff are shown in  Figs.~\ref{fig:wave_func_DDast_gs3.4} and \ref{fig:expvalue_DDast_gs3.4}, respectively.
As evidenced by the wave function plot in Fig.~\ref{fig:wave_func_DDast_gs3.4} and the probability in Table~\ref{tab:results_DDast}, the $DD^*(^3S_1)$ channel dominates, which is the lowest threshold channel. 
In Fig~\ref{fig:expvalue_DDast_gs3.4},  it is found that the 
mixing between the $DD^*(S)$ and $DD^*(D)$ 
channels driven by the tensor force
plays a significant role in producing the bound state. Furthermore, the mixing between $DD^*(S)$ and $D^*D^*(S)$, as well as $D^*D^*(D)$, also contributes to the binding. 
These results indicate 
that the $D^*D^*$ channel should be included 
in the $DD^\ast$ molecular state,
because of HQSS and the $DD^*$ mixing effect.

\begin{figure}[htbp] 
\centering
\includegraphics[width=0.9\linewidth,clip]{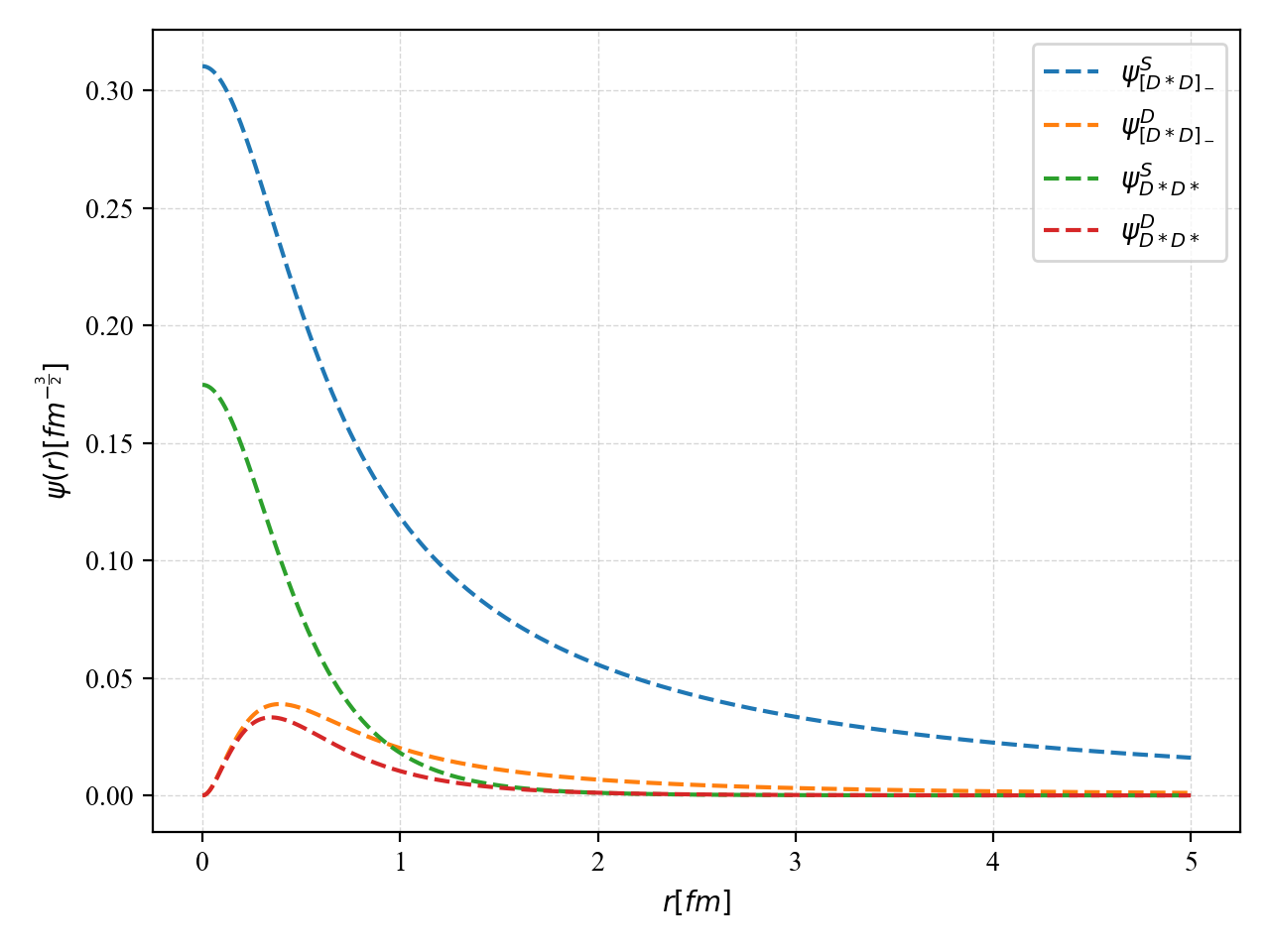}
\caption{Wave functions of the $D^{(*)}D^{(*)}$ molecule with $g_\sigma = g_\sigma^{(2)}=0.76$. The value $\psi(r)$ is given in unit of $[\rm{fm}^{-\frac{3}{2}}]$. The same notation as Fig.\ref{fig:wave_func_DDast_gs3.4} is used.}
\label{fig:wavefunction_DDast_gsig076}
\end{figure}

\begin{figure}[htbp]
\centering
\includegraphics[width=\linewidth,clip]{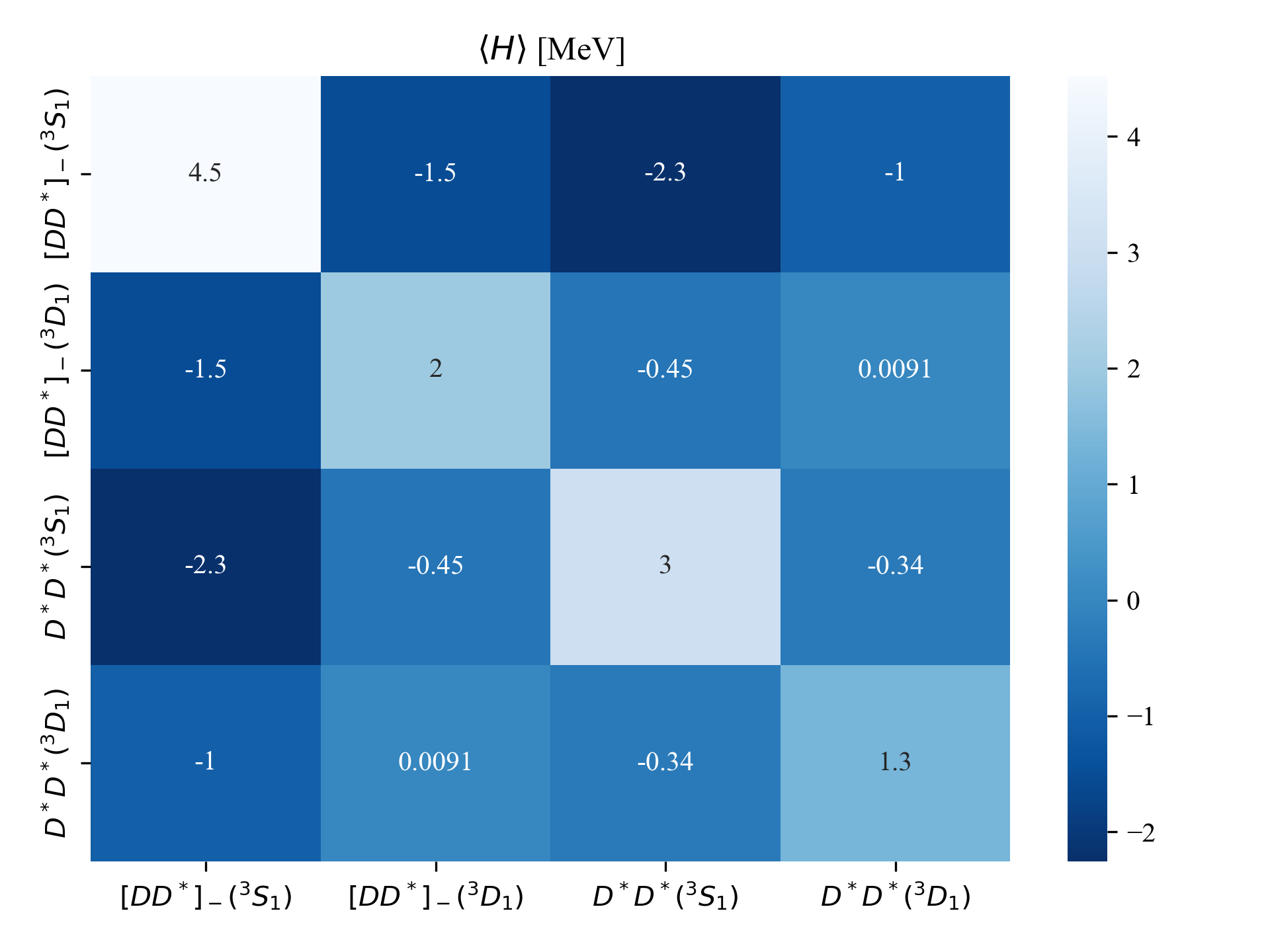} 
\caption{Energy expectation values $\langle H \rangle$ [MeV] of the $D^{(*)}D^{(*)}$ molecule with $g_\sigma = g_\sigma^{(2)}=0.76$. The same notation as Figure \ref{fig:expvalue_DDast_gs3.4} is used.} 
\label{fig:expectationvalue_DDast_gsig076}
\end{figure}

Actually, the coupling constant of $\sigma$ meson is ambiguous and 
the small value of 
$g_\sigma^{(2)} = 0.76$~\cite{Liu2008} has also been used in the literature.
We consider the case with the $\sigma$
coupling and
perform the coupled-channel analysis again. We obtain the cutoff $\Lambda=1631$ MeV to reproduce the $T_{cc}$ binding energy. 
In this model, the larger cutoff $\Lambda$ is needed to reproduce the $T_{cc}$ data due to the weaker $\sigma$ meson exchange.
However, we obtain that the qualitative properties of the bound state are the same as those of the case with $g_{\sigma}^{(1)}=3.4$.
The bound state properties, wave functions and energy expectation values are summarized in Table~\ref{tab:results_DDast_gsig076} and Figs.~\ref{fig:wavefunction_DDast_gsig076} and \ref{fig:expectationvalue_DDast_gsig076}, respectively.

\begin{table}[htbp]
    \centering
    \caption{
            Bound state properties of the $D^{(*)}D^{(*)}$ molecule with various cutoff $\Lambda$ with $g_\sigma = g_\sigma^{(2)}=0.76$.
            The cutoff $\Lambda=1631$ MeV is determined to reproduce the $T_{cc}$ binding energy.
            The same notation as Table~\ref{tab:results_DDast} is used.
            }
    \label{tab:results_DDast_gsig076}    
    \begin{ruledtabular}    
    \begin{tabular}{lccc}
        $\Lambda$ [MeV]& 1610 & 1631 & 1650 \\
    \hline 
        $\mathrm{B_{in}}$ [MeV]& 0.053 & 0.271 & 0.494 \\
    $P_{[DD^*]_{-}}(^3S_1)$ & 0.988 & 0.975 & 0.966 \\
    $P_{[DD^*]_{-}}(^3D_1)$ & 0.00545 & 0.0108 & 0.0132\\
    $P_{D^*D^*}(^3S_1)$ & 0.00527 & 0.0125 & 0.0175\\
    $P_{D^*D^*}(^3D_1)$ & 0.000941 & 0.00209 & 0.00285 \\
    $\sqrt{\langle r^2 \rangle}$ [fm]& 12.57 & 6.39 & 4.83\\  
    \end{tabular}
    \end{ruledtabular}    
\end{table}

\section{$\bar{D}\Xi_{cc}$ molecule}
\label{sec:DXiccmolecule}

In the previous section, we study the $T_{cc}$ state as a $D^{(\ast)}D^{(\ast)}$ molecule. With respect to superflavor symmetry~\cite{georgi1990,savage1990}, we expect that there should be a $\bar{D}^{(\ast)}\Xi_{cc}^{(\ast)}$ state that is a partner of $D^{(\ast)}D^{(\ast)}$ by replacing a charm quark as $\bar{c}\bar{c}$ diquark\footnote{We note that the superflavor partner of $D^{(\ast)}D^{(\ast)}$ should be $D^{(\ast)}\bar{\Xi}_{cc}^{(\ast)}$. In this paper, however, we study $\bar{D}^{(\ast)}\Xi_{cc}^{(\ast)}$ with $C=+1$ rather than the anti-charm state $D^{(\ast)}\bar{\Xi}_{cc}^{(\ast)}$. These states are equivalent each other by charge conjugation.} 
We study a coupled-channel analysis of $\bar{D}^{(*)}\Xi_{cc}^{(*)}$ molecular state, where with respect to HQSS, the mixing of the HQS doublet states, $\bar{D}$ and $\bar{D}^\ast$ for mesons and $\Xi_{cc}$ and $\Xi_{cc}^\ast$ for baryons,
are introduced.
Thus, we
consider the following coupled channels:
\begin{align}
\begin{pmatrix}
\bar{D}\Xi_{cc} (^2S_{\frac{1}{2}}) \\
\bar{D}\Xi_{cc}^* (^4D_{\frac{1}{2}}) \\
\bar{D}^*\Xi_{cc} (^2S_{\frac{1}{2}}, ^4D_{\frac{1}{2}})  \\
\bar{D}^*\Xi_{cc}^* (^2S_{\frac{1}{2}}, ^4D_{\frac{1}{2}}, ^6D_{\frac{1}{2}})
\end{pmatrix}    
,   
\end{align}
including possible mixing between
$S$ and $D$ wave states 
by the tensor force. 
The $\bar{D}^{(*)}\Xi_{cc}^{(*)}$ system has more channels than the $D^{(*)}D^{(*)}$ one shown in Eq.~\eqref{eq:DD*channels}. 

We define a doubly heavy baryon doublet field $\psi^\mu_{QQ}$ which is a linear combination of double heavy baryons $\mathcal{B}_{QQ}(\frac{1}{2}^+)$ and $\mathcal{B}^{*\mu}_{QQ}(\frac{3}{2}^+)$ degenerating in HQ limit as follows~\cite{Meng:2022ozq}
\begin{align}
\psi_{QQ}^{\mu} &= \mathcal{B}^{*\mu}_{QQ} + \sqrt{\frac{1}{3}}(\gamma^\mu + v^\mu) \gamma^5 \mathcal{B}_{QQ} ,\\
\bar{\psi}_{QQ}^{\mu} &= \bar{\mathcal{B}}^{*\mu}_{QQ} - \sqrt{\frac{1}{3}}  \bar{\mathcal{B}}_{QQ} \gamma^5 (\gamma^\mu + v^\mu) ,
\end{align}
where $\mathcal{B}_{QQ} = \Lambda_+ B_{QQ}$, and $\mathcal{B}^{*\mu}_{QQ} = \Lambda_+ B^{*\mu}_{QQ}$ with 
the positive-energy projection operator $\Lambda_+ = \frac{1 + \Slash{v}}{2}$.

$B_{QQ}$ is Dirac spinor, while $B^{*\mu}_{QQ}$ is Rarita-Schwinger spinor. 
In the ground state of doubly heavy baryons, the spin of the heavy diquark $QQ$ must be $1$ due to the Fermi-Dirac  
statistics. Therefore, the total spin of the doubly heavy baryon is given by 
combination of the $QQ$ spin 1 and the rest light quark spin $\frac{1}{2}$, i.e., $1 \otimes \frac{1}{2} = \frac{1}{2} \oplus \frac{3}{2}$. These two spin states are degenerate in HQ limit, and correspond to $\Xi_{cc}$ and $\Xi_{cc}^{*}$ in the charm sector. 
We utilize the experimental value of  the $\Xi_{cc}$ mass obtained by the LHCb collaboration~\cite{aaij2017}, while  the mass of $\Xi_{cc}^*$ with no data is estimated by
using the quark model mass relation, $\frac{3}{4} (m_{\Xi_{cc}^*} - m_{\Xi_{cc}}) = m_{D^*} -m_D$~\cite{Brambilla:2005yk,Fleming:2005pd,Hu:2005gf}. This yields $m_{\Xi_{cc}} \simeq 3727$ MeV. 

We employ the effective Lagrangians satisfying
HQSS and $SU(3)$ flavour symmetry. The interaction Lagrangians
in the leading order of  the $1/m_Q$ expansion 
are as follows.
The interaction Lagrangians
for the pseudoscalar meson and the scalar $\sigma$ meson are 
given by 
\begin{align}
\mathcal{L}_{\psi_{QQ}\psi_{QQ}A } =& \hat{g}\bar{\psi}_{QQ}^{\mu} A^{\nu} \gamma_{\nu} \gamma^5 \psi_{QQ \mu} \notag \\
=& \hat{g} \bar{\mathcal{B}}^{*\mu}_{QQ} A^\nu \gamma_\nu \gamma^5 \mathcal{B}^{*}_{QQ\mu} \notag \\ 
&- 2\sqrt{\frac{1}{3}} \hat{g} \left( \bar{\mathcal{B}}^{*\mu}_{QQ} A_\mu \mathcal{B}_{QQ} + \bar{\mathcal{B}}_{QQ} A_\mu \mathcal{B}^*_{QQ\mu} \right) \notag \\ 
&+ \frac{1}{3} \hat{g} \bar{\mathcal{B}}_{QQ} A^\nu \gamma_\nu \gamma^5 \mathcal{B}_{QQ} ,
\label{eq:L_Xipi} \\
\mathcal{L}_{\psi_{QQ}\psi_{QQ}\sigma} 
=& -\hat{g}_\sigma \bar{\psi}_{QQ}^{\mu} \sigma \psi_{QQ \mu} \notag \\
=& \hat{g}_{\sigma}\bar{\mathcal{B}}_{QQ} \sigma \mathcal{B}_{QQ} - \hat{g}_{\sigma} \bar{\mathcal{B}}^{* \mu}_{QQ} \sigma \mathcal{B}^*_{QQ \mu} \ .
\label{eq:L_Xisig}
\end{align}
Here and hereafter, we omit the isospin indices for notational simplicity. 
The interaction
Lagrangians for the vector mesons 
are 
\begin{align}
\mathcal{L}_{HH\beta} 
= & \, i\hat{\beta} \bar{\psi}_{QQ}^{\mu} v_\nu \rho^\nu \psi_{QQ \mu} \notag \\ 
= & \, - \frac{\hat{\beta} \hat{g}_v}{\sqrt{2}} \bar{\mathcal{B}}^{* \mu}_{QQ} v_\nu \hat{\rho}^\nu \mathcal{B}^*_{QQ \mu} + \frac{\hat{\beta} \hat{g}_v}{\sqrt{2}} \bar{\mathcal{B}}_{QQ} v_\nu \hat{\rho}^\nu \mathcal{B}_{QQ} , 
\label{eq:L_Xiv_beta} \\
\mathcal{L}_{HH\lambda} = & \, \hat{\lambda} \bar{\psi}_{QQ}^{\mu} \sigma_{\alpha \beta} F^{\alpha \beta} \psi_{QQ \mu} \notag\\
= & \, \frac{ \hat{\lambda} \hat{g}_v}{\sqrt{2}} \left( \bar{\mathcal{B}}^{* \mu}_{QQ} \varepsilon_{\alpha \beta \gamma \delta}v^\alpha \gamma^\beta \gamma^5 \left( \partial^\gamma \hat{\rho}^\delta - \partial^\delta \hat{\rho}^\gamma \right) \mathcal{B}^*_{QQ \mu} \right) \notag \\
&+ i 2 \sqrt{\frac{2}{3}} \hat{\lambda} \hat{g}_V \bar{\mathcal{B}}^{*}_{QQ \alpha} \gamma_\beta \gamma^5 \left( \partial^\alpha \hat{\rho}^\beta - \partial^\beta \hat{\rho}^\alpha \right) \mathcal{B}_{QQ} \notag \\
&+ i 2 \sqrt{\frac{2}{3}} \hat{\lambda} \hat{g}_V \bar{\mathcal{B}}_{QQ} \gamma_\alpha \gamma^5 \left( \partial^\alpha \hat{\rho}^\beta - \partial^\beta \hat{\rho}^\alpha \right) \mathcal{B}^*_{QQ \beta} \notag \\
& -\frac{i}{3\sqrt{2}} \hat{\lambda} \bar{\mathcal{B}}_{QQ} \gamma_\alpha \gamma_\beta (\partial^\alpha \hat{\rho}^\beta - \partial^\beta \hat{\rho}^\alpha) \mathcal{B}_{QQ} \ .
\label{eq:L_Xiv_lam}
\end{align}
The values of the coupling constants between $\Xi_{cc}$ and $\pi, \sigma, \rho, \omega$ are unknown due to a lack of experimental data. Therefore, we estimate the coupling constants 
based on the superflavor symmetry by using those of $\bar{D}$ as a reference. 
Since the quark configuration of $\bar{D}^{(*)}$ and $\Xi_{cc}^{(*)}$ is $\bar{Q}q$ and $QQq$, 
respectively, the heavy quark objects should have the same color source represented by the color anti triplet to make a color singlet hadron, 
i.e. $\bar{Q} \sim \bar{3}$ representation and  $QQ \sim \bar{3}$
(we note $Q \otimes Q \sim \bar{3} \otimes \bar{3} = \bar{3} \oplus 6$).  
The heavy diquark is a sizable object with finite mass, but it becomes a point-like particle in
the HQ limit. Thus, the heavy diquark can be identified with anti-heavy quark as a point-like static colour source in the HQ limit. Thus we obtain the coupling constants of $\Xi_{cc}$ to light hadrons 
from the $\bar{D}$ coupling constants by the superflavor symmetry. 
In OBEM, only the light quark is involved in the interaction of boson exchange and the light quark structures in $\bar{D}$ and $\Xi_{cc}^*$ are the same in the HQ limit. 
Thus the coupling constants of $\Xi_{cc}$ are given by those of $\bar{D}$ as 
\begin{align}
    \hat{g} &= 2g , \label{eq:g_hat}\\
    \hat{g}_\sigma &= 2g_\sigma ,\\
    \hat{\beta} &= 2\beta ,\\
    \hat{\lambda} &= 2\lambda .  \label{eq:lambda_hat}
\end{align}
We note that the factor of 2 is from
the normalization factor~\cite{hu2006}.
We also use the same cutoff parameter obtained in the $D^{(\ast)}D^{(\ast)}$ study fitted to reproduce the $T_{cc}$ data.

\begin{table}[htbp]
    \centering
    \caption{Bound state properties 
            of the $\bar{D}^{(*)}\Xi_{cc}^{(*)}$ molecule
            with various cutoff $\Lambda$ and $g_{\sigma}=g_\sigma^{(1)}=3.4$. 
            The binding energy $\mathrm{B_{in}}$ is measured from the $\bar{D}\Xi_{cc}$ threshold.
            The cutoff $\Lambda=1182$ MeV is determined from the $T_{cc}$ binding energy in Sec.~\ref{sec:DD*molecule}.
            The same notation as Table~\ref{tab:results_DDast} is used.
        }
    \label{tab:results_DbarXi}    
    \begin{ruledtabular}        
    \begin{tabular}{lcccc}
    $\Lambda$ [{MeV}] & 1160 & {1182} & 1200\\
\hline 
    $\mathrm{B_{in}}$  \, [{MeV}] & 5.65 & {7.46} & 9.10\\
    $P_{\bar{D}\Xi_{cc}} (^2S_{\frac{1}{2}})$ & 0.99 & {0.98} & 0.98\\
    $P_{\bar{D}\Xi_{cc}^*} (^4D_{\frac{1}{2}})$ & 0.000008 & {0.000012} & 0.000017\\
    $P_{\bar{D}^*\Xi_{cc}} (^2S_{\frac{1}{2}})$ & 0.00056 & {0.00076} & 0.00095\\
    $P_{\bar{D}^*\Xi_{cc}} (^4D_{\frac{1}{2}})$ & 0.0026 & {0.0030} & 0.0033\\
    $P_{\bar{D}^*\Xi_{cc}^*} (^2S_{\frac{1}{2}})$ & 0.0021 & {0.0028} & 0.0035 \\
    $P_{\bar{D}^*\Xi_{cc}^*} (^4D_{\frac{1}{2}})$ & 0.00068 & {0.00074} & 0.00079 \\
    $P_{\bar{D}^*\Xi_{cc}^*} (^6D_{\frac{1}{2}})$ & 0.0085 & {0.0097} & 0.011\\
    $\sqrt{\left\langle r^2\right\rangle}$  [fm] & 1.54 & {1.38} & 1.28\\
    \end{tabular}
    \end{ruledtabular}    
\end{table}

\begin{figure}[htbp] 
\centering
\includegraphics[width=0.9\linewidth,clip]{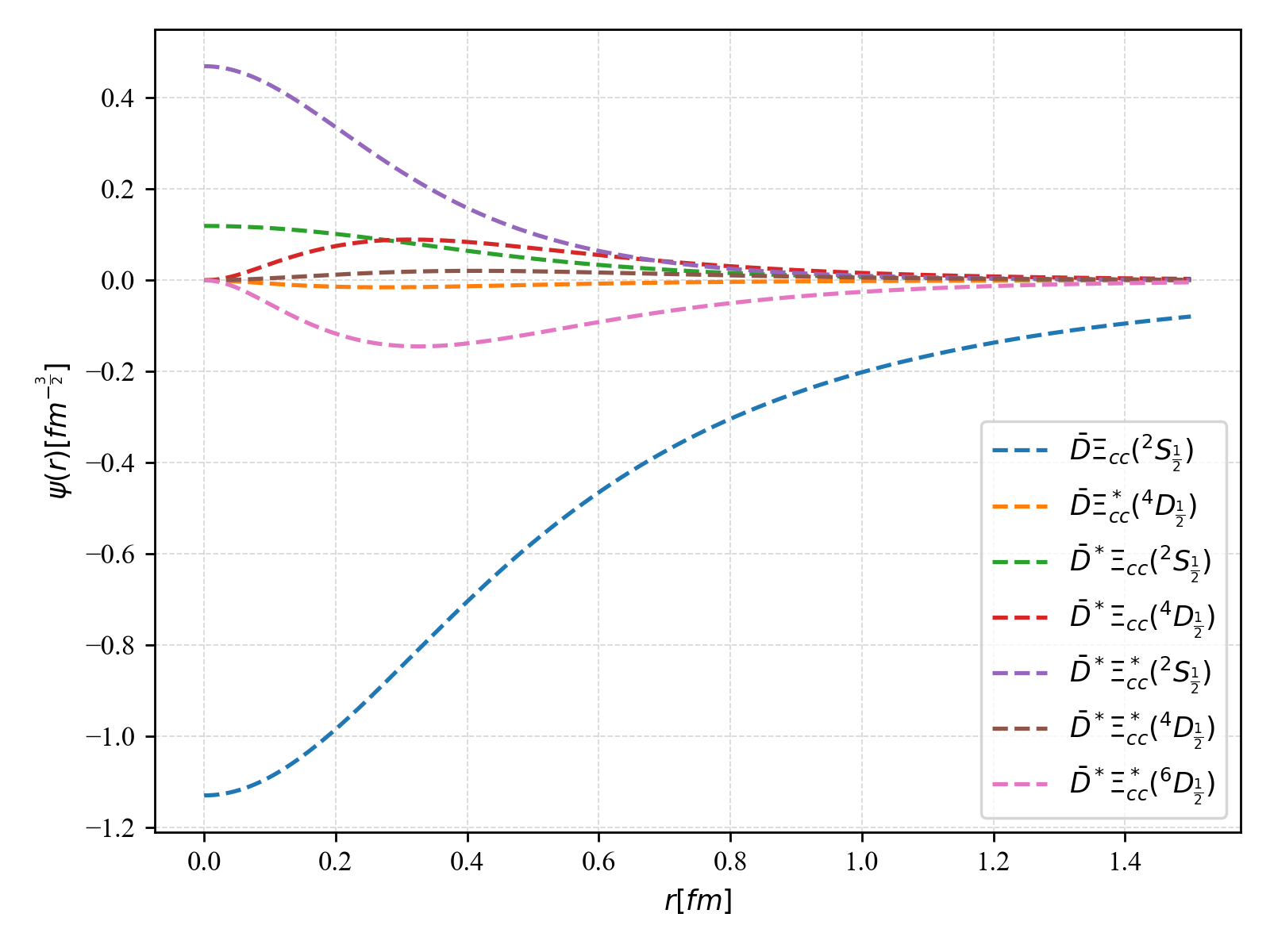}
\caption{Wave functions of the $\bar{D}^{(*)}\Xi_{cc}^{(*)}$ molecule with $g_\sigma=g_\sigma^{(1)}=3.4$. The value $\psi(r)$ is given in unit of $[\rm{fm}^{-\frac{3}{2}}]$.}
\label{fig:results_DbarXi_wavefunc}  
\end{figure}

\begin{figure}[htbp] 
\centering
\includegraphics[width=\linewidth,clip]{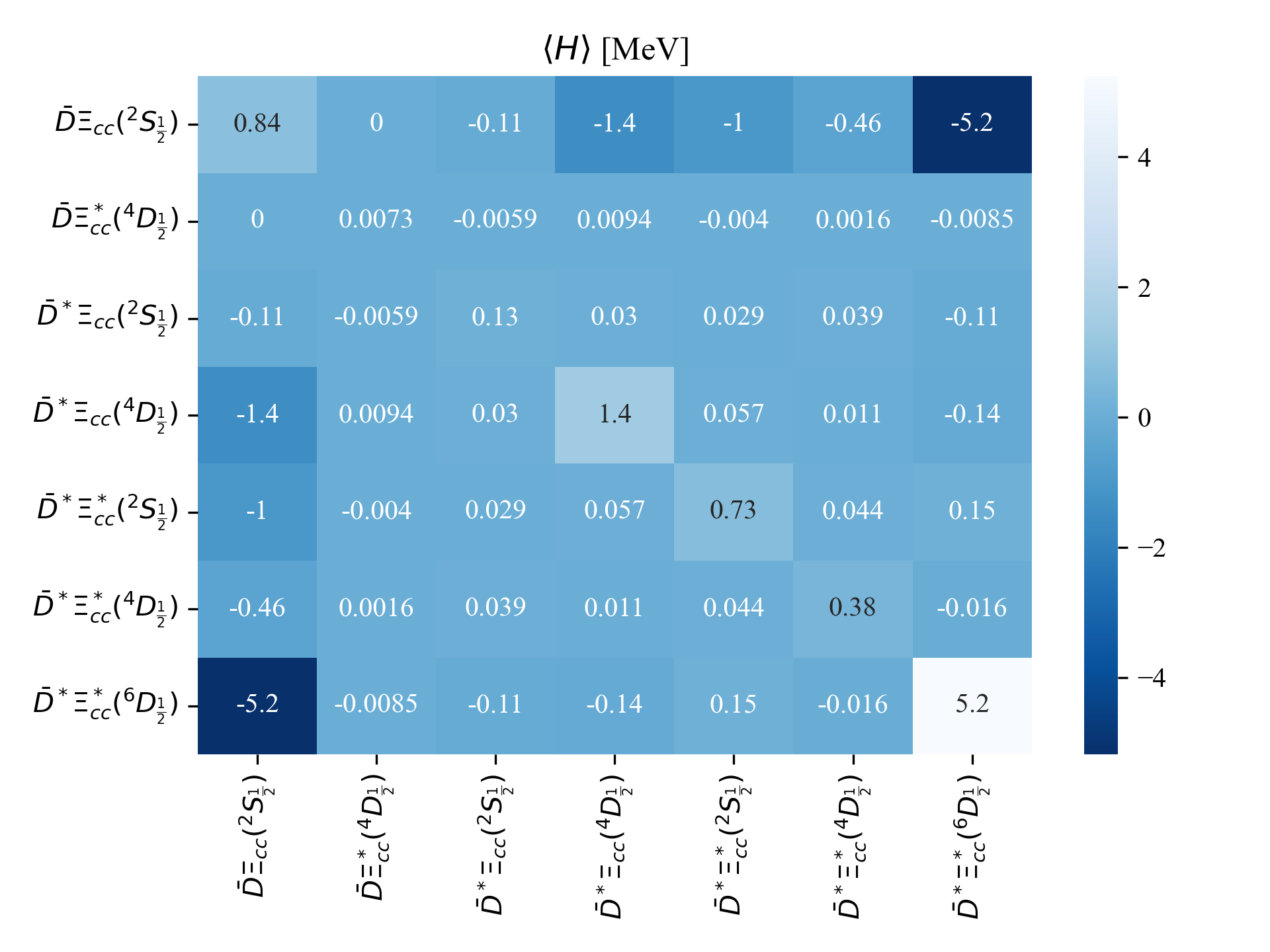}
\caption{The energy expectation values $\langle H \rangle$ [MeV] of the $\bar{D}^{(*)}\Xi_{cc}^{(*)}$ molecule with $g_\sigma=g_\sigma^{(1)}=3.4$. The same notation as Fig.~\ref{fig:expvalue_DDast_gs3.4} is used.}
\label{fig:results_DbarXi_H}  
\end{figure}

By solving coupled-channel Schr\"odinger equations using 
Gaussian expansion method,
we obtain a bound state of $\bar{D}\Xi_{cc}$ for $I(J^P)=0(1/2^-)$. The numerical results with $g_\sigma=g_\sigma^{(1)}=3.4$ are summarized in Table~\ref{tab:results_DbarXi}. 
The radial wave function and the matrix of the hamiltonian expectation value are also shown in Figs.~\ref{fig:results_DbarXi_wavefunc} and \ref{fig:results_DbarXi_H}, respectively. 
The probabilities in Table~\ref{tab:results_DbarXi} show that the dominant component of the wave functions is in 
the $\bar{D}\Xi_{cc}(^2S_{\frac{1}{2}})$ channel with $P_{\bar{D}\Xi_{cc}}(^2S_{\frac{1}{2}})\sim 99 \%$, which is also shown in the plot of the wave functions in Fig.~\ref{fig:results_DbarXi_wavefunc}. Interestingly, the second dominant component is the $\bar{D}^*\Xi_{cc}^*(^6D_{\frac{1}{2}})$ channel with $P_{\bar{D}^*\Xi_{cc}^*}(^6D_{\frac{1}{2}})\sim 1 \%$, while the threshold energy is far from the $\bar{D}\Xi_{cc}$ channel and furthermore it is a $D$-wave state. 
The important contribution of the $\bar{D}^*\Xi_{cc}^*(^6D_{\frac{1}{2}})$ channel is indicated in
the matrix of the 
energy expectation value in Fig.~\ref{fig:results_DbarXi_H}. 
The off-diagonal component between the $\bar{D}\Xi_{cc}(^2S_{\frac{1}{2}}) - \bar{D}^*\Xi_{cc}^*(^6D_{\frac{1}{2}})$ channels  contributes to a strong attraction which is produced by the tensor force. 
It's turned out that this attraction plays an important role to generate the bound state. In contrast, for the $D^{(*)}D^{(*)}$ molecule, both the $DD^{*}(^3S_1)-DD^{*}(^3D_1)$ and $DD^{*}-D^* D^*$ channel couplings are important to generate the attraction.
The mixing effects shown in the $D^{(*)}D^{(*)}$ and $\bar{D}^{(*)}\Xi_{cc}^{(*)}$ molecules imply that HQSS should be respected in heavy hadronic molecular systems.
In the bound state analysis, we couldn't obtain a bound state in the other $J^P$ states.

\begin{figure}[htbp] 
\centering
\includegraphics[width=\linewidth,clip]{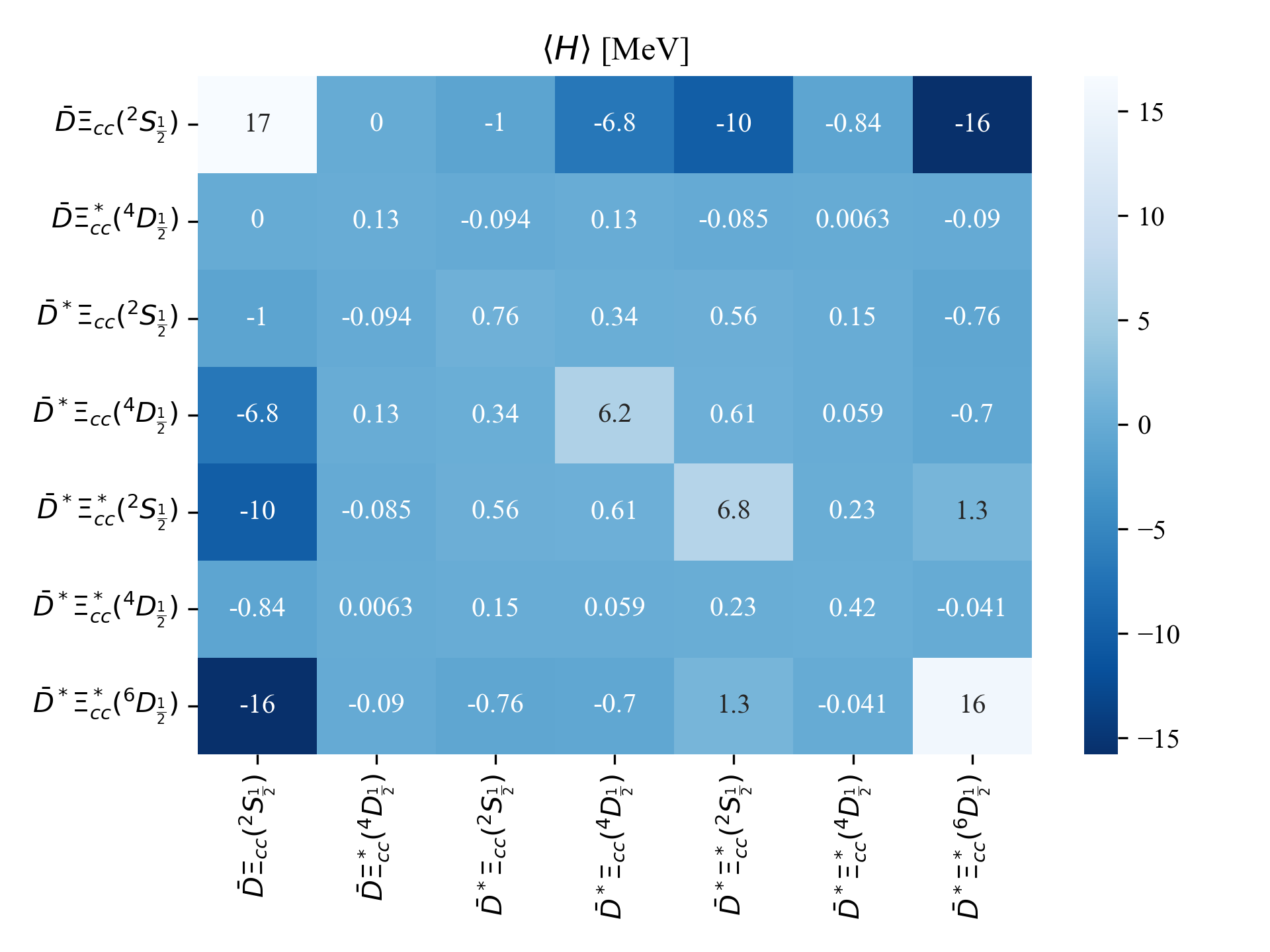}
\caption{Energy expectation values $\langle H \rangle$ [MeV] of the $\bar{D}\Xi_{cc}$ molecules with $g_{\sigma}=g_\sigma^{(2)}=0.76$. The same notation as Fig.~\ref{fig:expvalue_DDast_gs3.4} is used.}
\label{fig:results_DbarXi_H_gsig076}
\end{figure}

We also study the case with $g_\sigma = g_\sigma^{(2)}=0.76$. 
Although small sigma coupling is used, we find a bound state of $\bar{D}^{(*)}\Xi_{cc}^{(*)}$ for $I(J^P)=0(1/2^-)$.
The bound state properties are summarized in Table\ref{tab:results_DbarXi_gsig076}. In the probabilities, the dominant component is in
the $\bar{D}\Xi_{cc}(^2S_{\frac{1}{2}})$ channel, and the probability of the $\bar{D}^*\Xi_{cc}^*(^6D_{\frac{1}{2}})$ one induced by the tensor force is also negligible, $P_{\bar{D}^*\Xi_{cc}^*(^6D_{\frac{1}{2}})}\sim 2.3 \%$ as seen in the case with large $g_\sigma$. Interestingly the probability of the $\bar{D}^*\Xi_{cc}^*(^2S_{\frac{1}{2}})$ channel, which is suppressed in the case with $g_\sigma=g_\sigma^{(1)}=3.4$, is as large as that of the $\bar{D}^*\Xi_{cc}^*(^6D_{\frac{1}{2}})$ one. The $\bar{D}\Xi_{cc}(^2S_{\frac{1}{2}})-\bar{D}^*\Xi_{cc}^*(^2S_{\frac{1}{2}})$ mixing is induced by the spin-spin interaction of the vector meson exchanges. This contribution producing an attraction has an important role to generate the bound state as shown in the energy expectation values in Fig.~\ref{fig:results_DbarXi_H_gsig076}. The smaller strength of the $\sigma$ exchange potential emphasizes the role of the vector meson exchanges in addition to the $\pi$ exchange.

\begin{table}[htbp]
    \centering
    \caption{
        Bound state properties
        of the $\bar{D}^{(*)}\Xi_{cc}^{(*)}$ molecules with $g_{\sigma}=g_\sigma^{(2)}=0.76$. 
        The cutoff $\Lambda=1631$ MeV is determined from the $T_{cc}$ binding energy in Sec.~\ref{sec:DD*molecule}.
        The same        
        notation as Table~\ref{tab:results_DDast} is used.}
    \label{tab:results_DbarXi_gsig076}    
    \begin{ruledtabular}        
    \begin{tabular}{lcccc}
    $\Lambda$ [MeV] & 1610 & {1631} & 1650\\
\hline $\mathrm{B_{in}}$ [MeV] & 16.8 & {19.0} & 21.2\\
$P_{\bar{D}\Xi_{cc}} (^2S_{\frac{1}{2}})$ & 0.94 & 0.94 & 0.93\\
$P_{\bar{D}\Xi_{cc}^*} (^4D_{\frac{1}{2}})$ & 0.00015 & {0.00016} & 0.00018\\
$P_{\bar{D}^*\Xi_{cc}} (^2S_{\frac{1}{2}})$ & 0.0041 & {0.0044} & 0.0046\\
$P_{\bar{D}^*\Xi_{cc}} (^4D_{\frac{1}{2}})$ & 0.0083 & {0.0088} & 0.0092\\
$P_{\bar{D}^*\Xi_{cc}^*} (^2S_{\frac{1}{2}})$ & 0.024 & {0.027} & 0.030 \\
$P_{\bar{D}^*\Xi_{cc}^*} (^4D_{\frac{1}{2}})$ & 0.00081 & {0.00080} & 0.00078 \\
$P_{\bar{D}^*\Xi_{cc}^*} (^6D_{\frac{1}{2}})$ & 0.022 & {0.023} & 0.024\\
$\sqrt{\left\langle r^2\right\rangle}$ [fm] & 0.96 & 0.91 & 0.88\\       
    \end{tabular}
    \end{ruledtabular}    
\end{table}

\begin{figure}[htbp] 
\centering
\includegraphics[width=0.9\linewidth,clip]{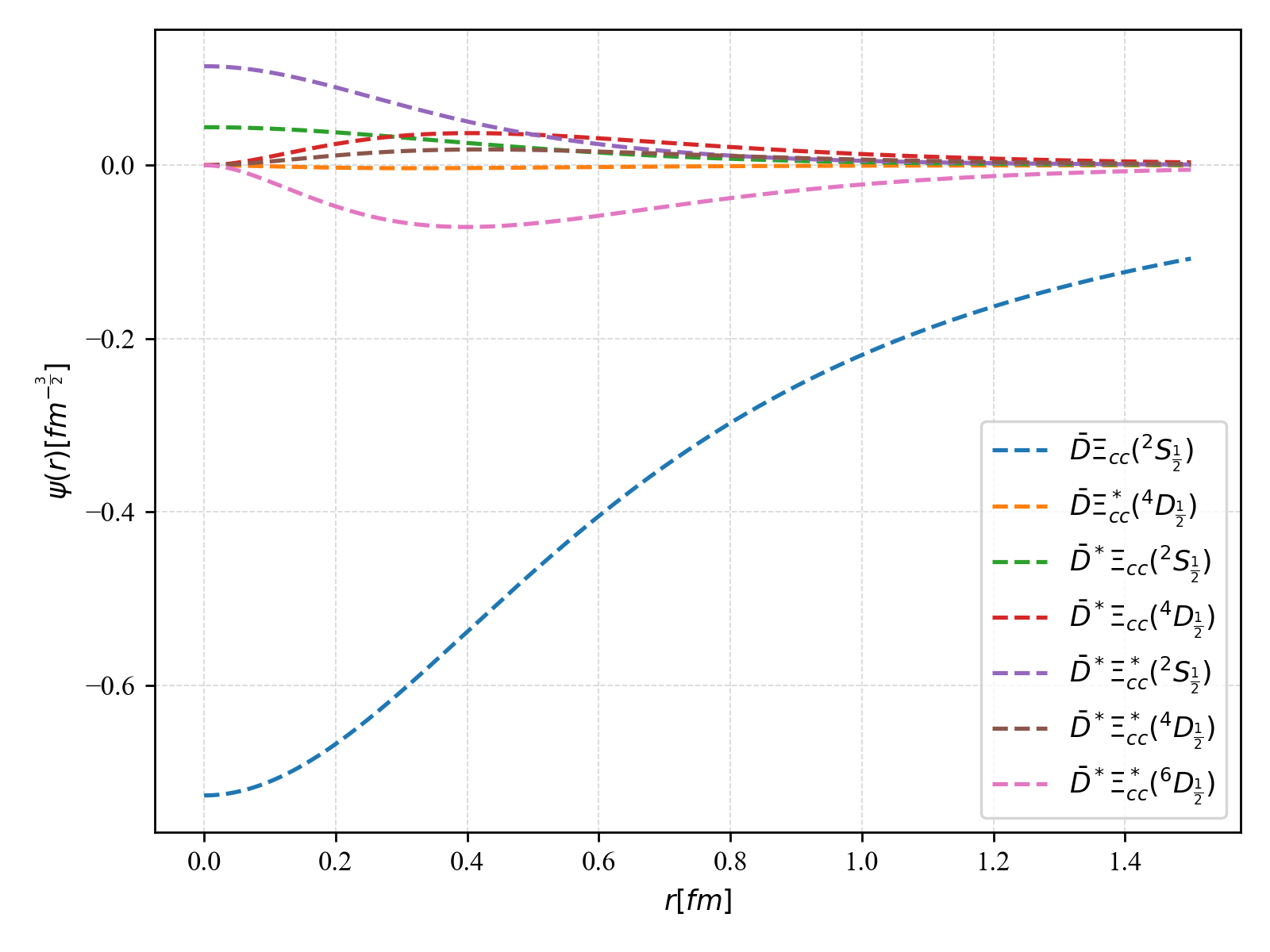} 
\caption{Wave functions of the $\bar{D}\Xi_{cc}$ molecules with $g_{\sigma}=g_\sigma^{(2)}=0.76$. The value $\psi(r)$ is given in unit of $[\rm{fm}^{-\frac{3}{2}}]$.  }
\label{fig:results_DbarXi_wavefunc_gsig076}
\end{figure}

The $\bar{D}^{(*)}\Xi_{cc}^{(*)}$ bound state is predicted under reasonable assumptions that $T_{cc}$ near the threshold has a $D^{(*)}D^{(*)}$ molecular structure and the superflavor symmetry is present between $\bar{D}$ mesons with a $\bar{c}$ quark and $\Xi_{cc}$ baryons with a $cc$ diquark. We expect that the $\bar{D}^{(*)}\Xi_{cc}^{(*)}$ state will be found in future experimental studies.

\begin{figure}[htbp] 
\centering
\includegraphics[width=0.9\linewidth,clip]{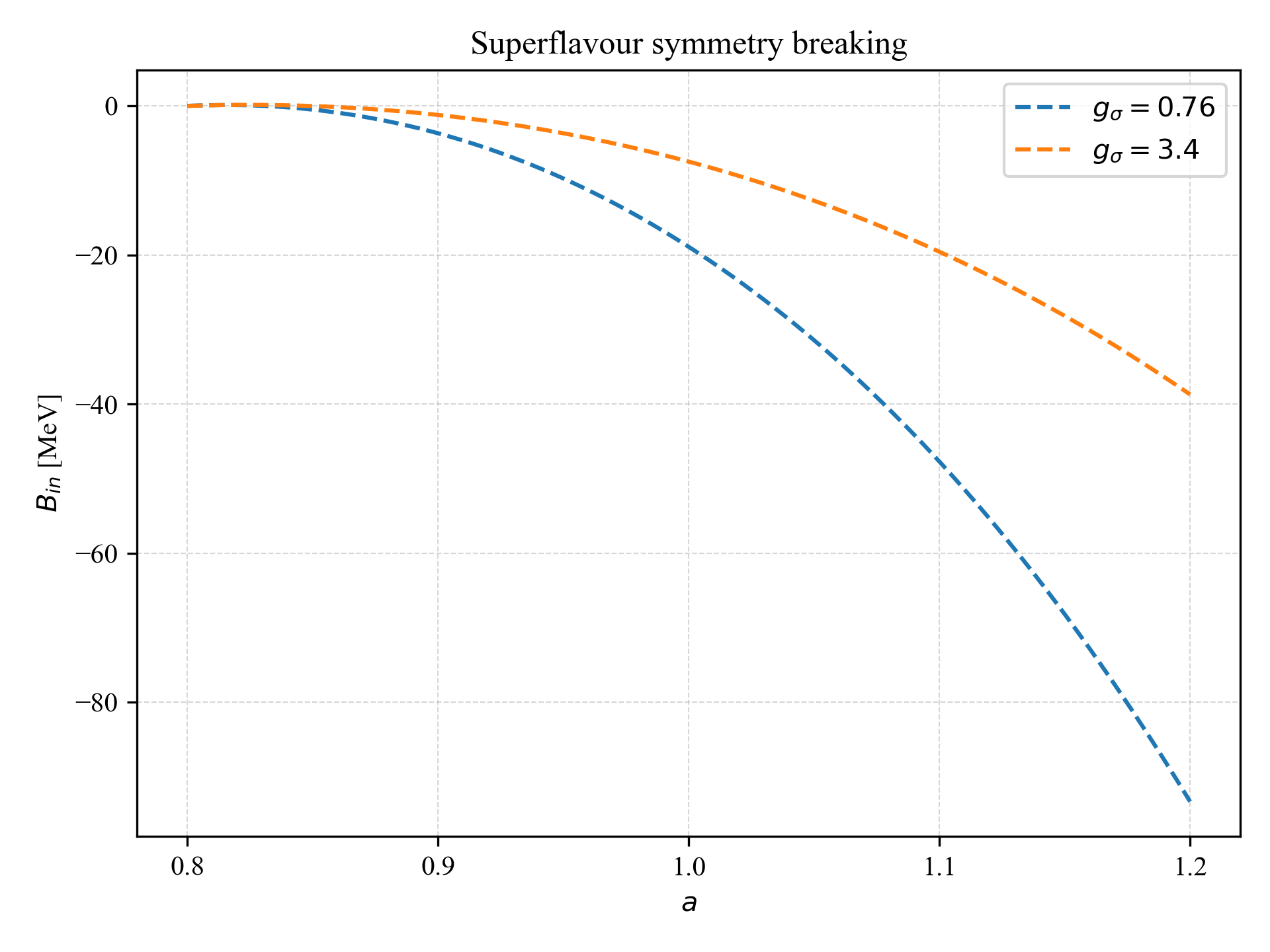}
\caption{\label{fig:Superflavour symmetry breaking} 
        Dependence of the $\bar{D}^{(*)}\Xi_{cc}^{(*)}$ binding energy ($B_{in}$ in unit of MeV) on the coupling constants with $g_\sigma=g_\sigma^{(1)}=3.4$ and $g_\sigma=g_\sigma^{(2)}=0.76$.
        }
\end{figure}

We obtain the coupling constants of $\Xi_{cc}^{(*)}$ by applying the superflavor symmetry between $\bar{D}^{(*)}$ and $\Xi_{cc}^{(*)}$. However, 
the error of the superflavor symmetry is estimated as $\Lambda_{QCD}/m_c v \sim 30 \% - 40 \%$ with $v$ being the velocity of the charm quark~\cite{savage1990,guo2013}, which is larger than the error $\Lambda_{QCD}/m_c \sim 10\%-15\%$ for the HQSS.
Therefore, we investigate the breaking effect of the superflavor symmetry
by evaluating the dependence of binding energies on the coupling constants. 
Here we multiply a factor $a$ to the coupling constants $\hat{g}$, $\hat{g}_\sigma$, $\hat{\beta}$ and $\hat{\lambda}$ in  Eqs.~\eqref{eq:g_hat}-\eqref{eq:lambda_hat} as
\begin{align}
 \hat{g} = 2a g \ , \quad  \hat{g}_\sigma = 2 a g_\sigma \ , \quad 
 \hat{\beta} = 2 a \beta \ , \quad  \hat{\lambda} = 2a \lambda \ ,
\end{align}
and vary $a$ from $0.8$ to $1.2$.
The results are shown in Fig.~\ref{fig:Superflavour symmetry breaking}

The dependence of the binding energy in coupling constants is not insignificant. When the ratio $a = 0.8$,
the bound state disappears in both models of $g_\sigma = g_\sigma^{(1)}= 3.4$ and $g_\sigma = g_\sigma^{(2)}=0.76$. However, the incorporation of such a significant error into the effective Lagrangian for doubly heavy baryons is not yet clearly established. 
Further theoretical and experimental studies are important to evaluate the the superflavor symmetry breaking effect in the effective theory. 

\section{Summary}
\label{sec:summary}

We have investigated the $D^{(\ast)}D^{(\ast)}$ and $\bar{D}^{(\ast)}\Xi_{cc}^{(\ast)}$ molecular states as candidates
of the heavy exotic hadrons.
The $D^{(\ast)}D^{(\ast)}$ molecule is considered to be a structure of the doubly charm tetraquark $T_{cc}$ near the threshold, reported by LHCb. We have studied a bound state of $D^{(\ast)}D^{(\ast)}$ with $I(J^P)=0(1^+)$, introducing the one boson exchange potential model. By fitting the cutoff parameter of the potentials, we obtained the model that reproduced the $T_{cc}$ binding energy. In this analysis, we have found that the $\sigma$ exchange potential produces the strong attraction. In addition, the off-diagonal terms of the potentials, especially given by the tensor force of the $\pi$ exchange and the spin-spin term of the vector meson exchange, also play the important role to make the bound state. The significant contributions from the off-diagonal components in the $DD^\ast-D^\ast D^\ast$ coupling indicate that the heavy quark spin symmetry inducing the small $DD^\ast$ mass splitting is important in a bound state of $D^{(\ast)}$ mesons.

We have also studied the $\bar{D}^{(\ast)}\Xi_{cc}^{(\ast)}$ bound state as a partner of the $D^{(\ast)}D^{(\ast)}$ bound state under the superflavor symmetry.  
The one boson exchange potential has also introduced as for the $\bar{D}^{(\ast)}\Xi_{cc}^{(\ast)}$ interaction, where unknown parameters for the $\Xi_{cc}^{(\ast)}$ baryons are determined by using the superflavor
symmetry from the parameters of the $\bar{D}^{(\ast)}$ mesons. Using the obtained potentials, we predicted the $\bar{D}^{(\ast)}\Xi_{cc}^{(\ast)}$ bound state with $I(J^P)=0(1/2^-)$. As discussed in the $D^{(\ast)}D^{(\ast)}$ bound state, we also found that the $\sigma$ exchange in the diagonal terms and the spin-dependent interactions in the off-diagonal terms play the important role to produce the attraction. 
The superflavor symmetry predicts the existence of the $\bar{D}^{(\ast)}\Xi_{cc}^{(\ast)}$ bound state, if the reported $T_{cc}$ is the $D^{(\ast)}D^{(\ast)}$ molecule. 
Our prediction will be useful to understand natures of the heavy hadron spectra including not only $\bar{D}^{(\ast)}\Xi_{cc}^{(\ast)}$ but also the $T_{cc}$ tetraquark and other hadronic molecules.

However, it is crucial to take into account the breaking of HQS caused by the finite charm quark mass. Therefore, we need to consider the next leading order of $\mathcal{O}(1/{m_Q})$ in the effective Lagrangian. Although the effective Lagrangian in the next leading order of $\mathcal{O}(1/{m_Q})$ have been studied~\cite{yasui2014}, the parameters of coupling constants have not been determined yet. Moreover, the validation of the the superflavor symmetry and the extent of its breaking have not been investigated. Therefore, more experiment data of exotic state with heavy quarks are needed.

\section*{Acknowledgment}
This work is in part supported by Grants-in-Aid for
Scientific Research under Grant Numbers JP20K14478
(Y.Y.), 20K03927, 23H05439 (M.H.). 

\bibliography{bio}

\end{document}